# Effects of weather and policy intervention on COVID-19 infection in Ghana

Iddrisu Wahab Abdul[*,1], Peter Appiahene[2], and Justice A. Kessie[1]

[1]Department of Mathematics and Statistics, University of Energy and Natural Resources, Sunyani, Ghana

[2]Department of Computer Science and Informatics, University of Energy and Natural Resources, Sunyani, Ghana

April 28, 2020

**Abstract**

As the number of COVID-19 cases continues to surge and the disease continues to wreak more havoc globally, new revelations concerning the spread and transmission of the virus continue to emerge from research every day. Even though laboratory and epidemiological studies have demonstrated the effects of ambient temperature on the transmission and survival of coronaviruses, not much has been done on the effects of weather on the spread of COVID-19. This study investigates the effects of temperature, humidity, precipitation, wind speed and the specific government policy intervention of partial lockdown on the new cases of COVID-19 infection in Ghana. Daily data on confirmed cases of COVID-19 from March 13, 2020 to April 21, 2020 were obtained from the official website of Our World in Data (OWID) dedicated to COVID-19 while satellite climate data for the same period was obtained from the official website of the National Aeronautics and Space Administration's (NASA's) Prediction of Worldwide Energy Resources (POWER) project. Considering the nature of the data and the objectives of the study, a time series generalized linear model which allows for regressing on past observations of the response variable and covariates was used for model fitting. The results indicate significant effects of maximum temperature, relative humidity and precipitation in predicting new cases of the disease. Also, results of the intervention analysis indicate that the null hypothesis of no significant effect of the specific policy intervention of partial lockdown should be rejected (p-value=0.0164) at a 5% level of significance. These findings provide useful insights for policymakers and the public.

*Correspondence to: iddrisu.wahab@uenr.edu.gh

# 1 Introduction

COVID-19 continues to wreak more havoc globally [1], [2]. As the number of infections keep surging high, deaths continue to increase at no small rate, active cases continue to scare the world [3], and a possible vaccine invention seems to be several months (if not years) away from now (though we do not lack the expertise, the technology and funding for such an expedition), and its treatment the world is quite unsure, leaving us all in horrendous fear. This pandemic has touched everything and activity in this world and has sent the world experts in health and data science to work to find possible solutions to this pandemic, while the experts in economics and policy makers have concerned themselves with the alleviation of the impact of COVID-19 on their nations' economy, its people and their businesses.

Researchers have investigated the role of pre-existing conditions such as aging, hypertension, etc. on COVID-19 deaths [4], [5], environmental factors that influences COVID-19 deaths [5], [6] and the role of climate on COVID-19 related deaths and new confirmed cases [7], [8].

The relationships between climate and infectious diseases have long been established as seasonal variations and climate sensitivities are well known to explain some variability in many infectious diseases. Statistical, process-based, and landscape models are the key categories of models for the estimation of potential climate effects on infectious diseases. These three model categories tackle very different problems. Statistical models for instance, require that the relationship between the current geographic spread of the disease and actual localized climate-specific conditions be inferred empirically [9].

Hitherto, there have been the confirmations that ultraviolet light has a sterilizing effect, because the radiation damages the virus's genetic material and its ability to replicate [10], [11]. This was first put out by [10], whose findings informed us that temperature and irradiance of COVID-19 will decide the trajectory of the pandemic at warmer regions as well as whether rising temperatures will change direction and have consequences on public health policy. They also showed that case and death counts at higher temperatures (>14 °C) when aligned for the epidemic stage had significantly lower rates of growth. However, according to [8], in warm-weather locations such as Singapore, Malaysia, and Thailand, COVID-19 have also proven lethal and pose wider concerns on the environmental factors.

[12] studied the association between temperature and number of new infections in 122 cities in China and made some important revelations including the fact that temperature less than 3 °C was positively linked to newly confirmed COVID-19 cases. Another work done by [13] on the "role of temperature and humidity in the modulation of the doubling time of COVID-19 cases" affirms what was revealed by [12] and [10]. Their conclusion was that, temperature and humidity explain a total of 18% variability in the disease doubling time while the rest (82%) may be attributed to containment initiatives, general health policies, population growth, and transport or cultural aspects.

In this study, we investigate the role of climate and specific government policy of partial lockdown in determining the number of new infections or confirmed cases in Ghana and to predict the future occurrences of this outbreak in the country. [14]–[16] studied the effects of lockdown on the number of new confirmed cases and/or on deaths. All of which revealed that lockdown serves an important intervention for preventing and minimizing the impact of an outbreak. Searching through literature, it appears Ghana at the moment does not have a statistical model that could predict future occurrences of new confirmed cases of COVID-19. To that extent, this study is novel.

## 2 Materials and methods

### 2.1 Study area

The study covers the whole of Ghana. Figure 1 is a map of the study area indicating the cumulative confirmed cases in each region as of April 25, 2020.

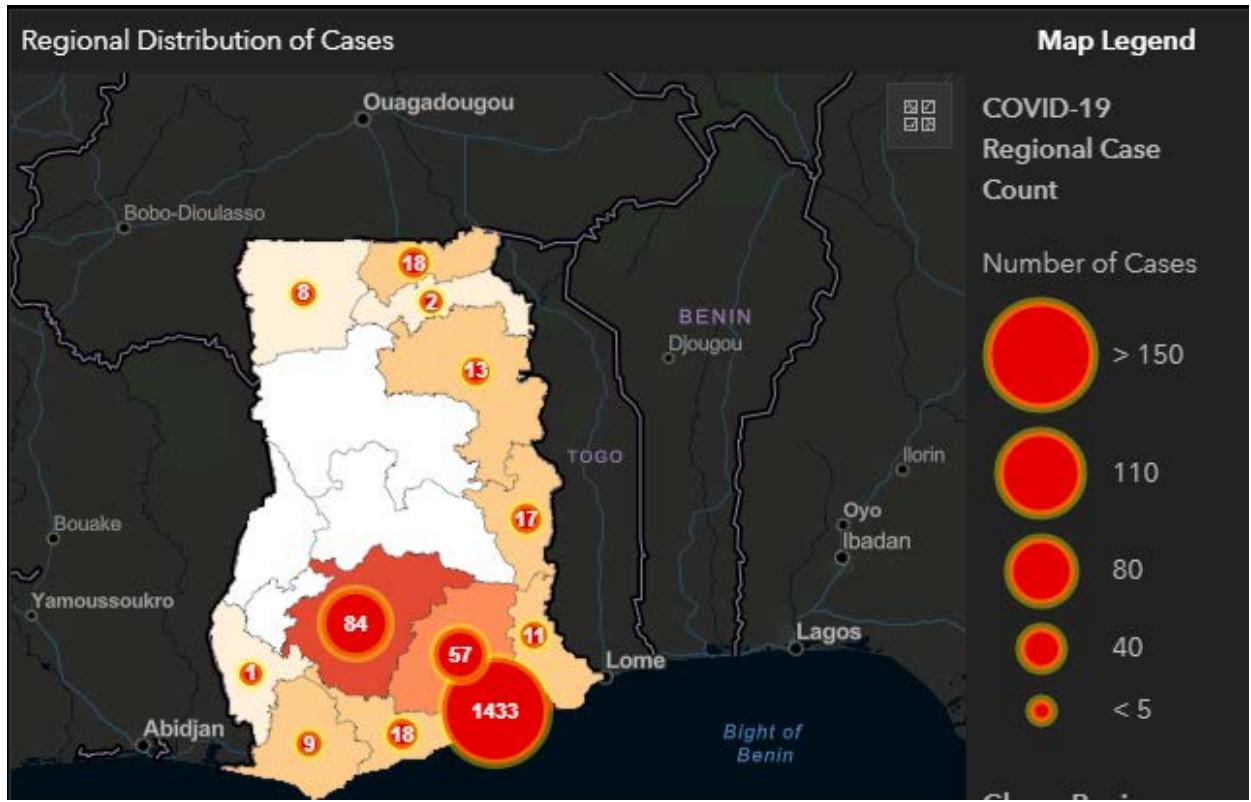

Figure 1: Map of study area (Source: https://www.ghanahealthservice.org/covid19/; accessed on April 28, 2020)

*2.2 Data*

Daily data on new cases of COVID-19 from March 13, 2020 to April 21, 2020 were obtained from the official website of Our World in Data (OWID) dedicated to COVID-19 (https://ourworldindata.org/coronavirus-source-data).

On the other hand, daily satellite climate data covering the same study period on maximum temperature, humidity, precipitation and maximum wind speed were obtained from the official website of the National Aeronautics and Space Administration's (NASA's) Prediction of Worldwide Energy Resources (POWER) project (https://power.larc.nasa.gov/data-access-viewer/).

*2.3 Statistical analysis*

To model the daily confirmed cases of COVID-19 in Ghana, which is a count variable, the generalized linear model (GLM) framework for time series of counts by [17] was adopted. These are models whose conditional mean of the response variable depends on previous observations of covariates and on its own previous values.

Denote the new cases of COVID-19 by $\{Y_t : t \in \mathbb{N}\}$, and denote by $\{X_t : t \in \mathbb{N}\}$ a time-varying $r$-dimensional covariate vector, say $\boldsymbol{X}_t = (X_{t,1}, \ldots, X_{t,r})^\top$. The conditional mean $E(Y_t | \mathcal{F}_{t-1})$ of the new cases is modeled by a process, say $\{\lambda_t : t \in \mathbb{N}\}$, such that $E(Y_t | \mathcal{F}_{t-1}) = \lambda_t$. Denote by $\mathcal{F}_t$ the history of the joint process $\{Y_t, \lambda_t, \boldsymbol{X}_{t+1} : t \in \mathbb{N}\}$ up to time $t$ including the covariate information at time $t + 1$. The general form of the model of interest developed by [17] is given in equation (1).

$$g(\lambda_t) = \beta_0 + \sum_{k=1}^{p} \beta_k \tilde{g}(Y_{t-i_k}) + \sum_{\ell=1}^{q} \alpha_\ell g(\lambda_{t-j_\ell}) + \boldsymbol{\eta}^\top \boldsymbol{X}_t, \qquad (1)$$

Where $g : \mathbb{R}^+ \to \mathbb{R}$ is a link function and $\tilde{g} : \mathbb{N}_0 \to \mathbb{R}$ is a transformational function. The parameter vector $\boldsymbol{\eta} = (\eta_1, \ldots, \eta_r)^\top$ corresponds to the effects of covariates. To allow for regression on arbitrary past observations of the response, define a set $P = \{i_1, i_2, \ldots, i_p\}$ and integers $0 < i_1 < i_2 \ldots < i_p < \infty$, with $p \in \mathbb{N}_0$. This enables regression on the lagged observations $Y_{t-i_1}, Y_{t-i_2}, \ldots, Y_{t-i_p}$. Parameters of the autocorrelation terms are represented by $\alpha_\ell$ and $\beta_k$.

The effect of government policy intervention such as the partial lockdown of the two largest cities in Ghana (Accra and Kumasi) and their environs on the daily confirmed cases of COVID-19 was analyzed using methods to test for such intervention effects developed by [18], [19]. These methods employ an approximate score test which has asymptotically a $\chi^2$ distribution, assuming some regularity conditions [19].

Data analysis and model fitting was done in R [20] using the "tscount" package [21].

# 3 Results

## 3.1 Descriptives

Descriptive statistics for the daily confirmed cases of COVID-19 and weather variables in Ghana are presented in Table 1. As of April 21, 2020 Ghana had recorded a total of 1,040 cases of COVID-19 and an average of about 26.72 new cases every day. Regarding the weather variables, average daily maximum temperature, precipitation, humidity and maximum wind speed were 34.4°C, 2.56mm, 68.99% and 3.33m/s respectively.

**Table 1: Descriptive statistics of daily confirmed cases of COVID-19 and satellite climate data in Ghana**

| Variables | N | Mean | Sd | Min | Max | Range | skew | Se |
|---|---|---|---|---|---|---|---|---|
| New_Cases | 39 | 26.72 | 51.55 | 0 | 208 | 208 | 2.38 | 8.25 |
| Max_Temp (°C) | 39 | 34.46 | 2.94 | 28.52 | 40.44 | 11.92 | 0.19 | 0.47 |
| Precipitation (mm) | 39 | 2.56 | 6.09 | 0.10 | 37.95 | 37.85 | 5.04 | 0.97 |
| Humidity (%) | 39 | 68.99 | 9.95 | 48.62 | 84.33 | 35.71 | -0.29 | 1.59 |
| Max_Wind (m/s) | 39 | 3.33 | 0.74 | 1.56 | 4.79 | 3.23 | -0.23 | 0.12 |

Figure 2 shows time series of daily new cases of COVID-19, maximum temperature, precipitation, humidity and maximum wind speed. The plot is an excellent way to begin comprehending the sort of process that produced the data and the relationships that exists therein. It is observed from the plot that neither the daily confirmed cases of COVID-19 nor the weather variables show any trend.

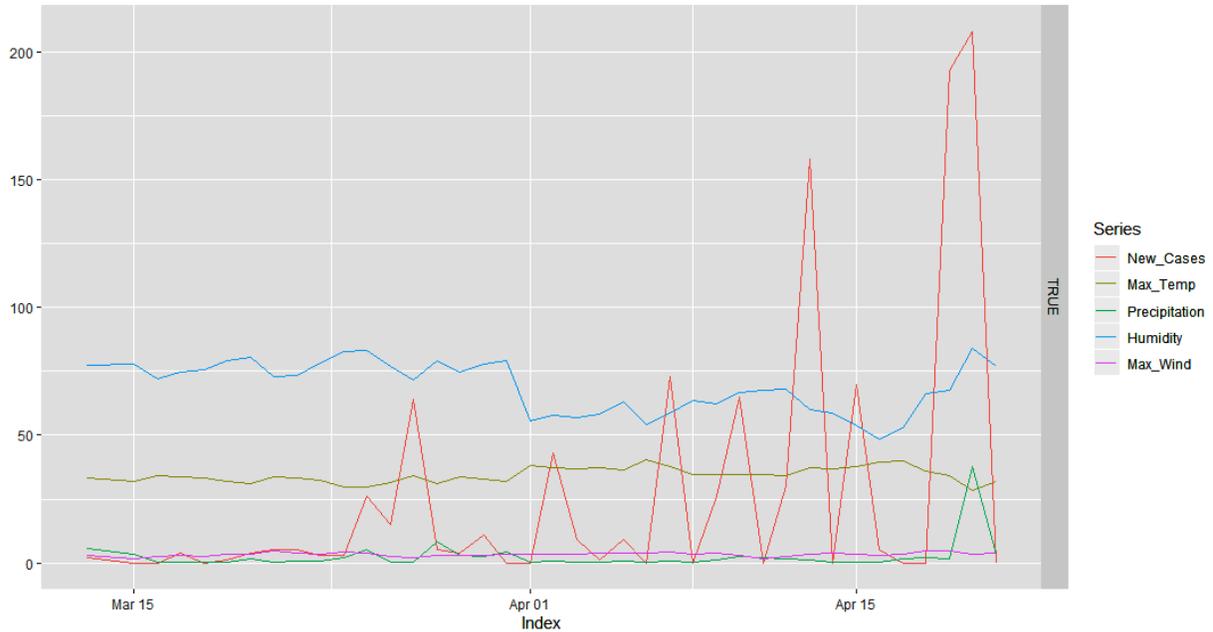

**Figure 2: Time series of daily confirmed cases of COVID-19, and weather variables from March 13, 2020 to April 21, 2020**

*3.2 Relationship between weather variables and new cases of COVID-19*

Correlations between the weather variables and confirmed cases of COVID-19 are shown in Figure 3. While the new cases of COVID-19 correlated negatively with maximum temperature and humidity, it rather correlated positively with precipitation and maximum wind. This means that increasing maximum temperature and humidity results in decreasing new cases of COVID-19 whereas increasing precipitation and maximum wind results in increasing new cases of COVID-19. Among the weather variables, weak correlations were observed.

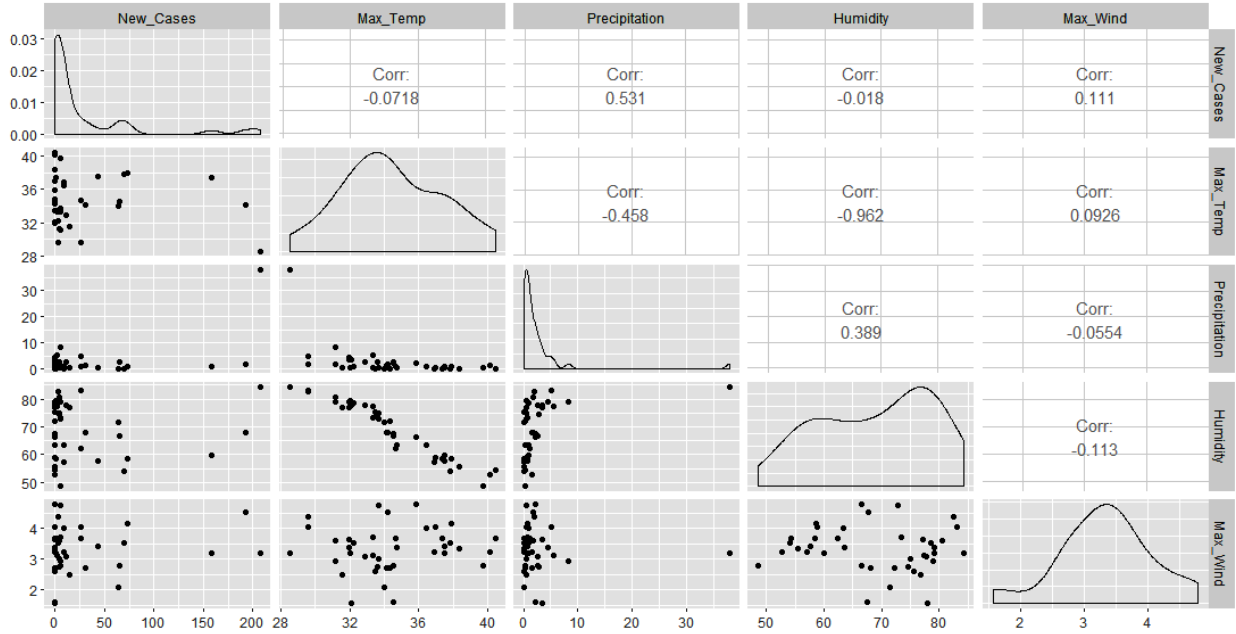

**Figure 3: Correlations between confirmed cases of COVID-19 and the weather variables**

Estimates of the model parameters are presented in Table 2. The log-linear model with the logarithmic link was chosen for model fitting because it allows for negative covariate effects. Also, in order to capture the short range serial dependence, a first order autoregressive term (beta_1) was included. Maximum temperature (Max_Temp), precipitation (Precipitation), relative humidity (Humidity), and maximum wind speed (Max_Wind) were included as explanatory variables. A deterministic covariate (linearTrend) was included to describe a linear trend.

The fitted model for the number of new cases of COVID-19 (New_Cases$_t$) in day $t$ is accordingly given by

$$\text{New\_Cases}_t | \mathcal{F}_{t-1} \sim Poisson(\lambda_t)$$

With

$$\log(\lambda_t) = 13.8430 - 0.2279 \text{New\_Cases}_{t-1} - 0.2267 \text{Max\_Temp}_t + 0.0407 \text{Precipitation}_t - 0.0605 \text{Humidity}_t - 0.0930 \text{Max\_Wind}_t + 28.7428 t/365$$

It is observed from Table 2 that the 95% confidence interval for the estimated coefficient corresponding to the first order autocorrelation (beta_1) does not contain zero, indicating a clear dependence of confirmed COVID-19 cases on the number of new cases of the preceding day. It is also observed that three of the weather variables considered (maximum temperature, precipitation and relative humidity) were significant predictors of the number of new cases of COVID-19. However, maximum wind speed was not a significant predictor of the confirmed cases of COVID-19. The linear trend was also found to be significant.

**Table 2: Estimates of model parameters**

|             | Estimate | Std.Error | 95% CI(lower) | 95% CI(upper) |
|---|---|---|---|---|
| (Intercept) | 13.8430 | 2.8291 | 8.7491 | 19.9134 |
| beta_1 | -0.2279 | 0.0884 | -0.4334 | -0.0968 |
| Max_Temp | -0.2267 | 0.0519 | -0.3377 | -0.1326 |
| Precipitation | 0.0407 | 0.0056 | 0.0288 | 0.051 |
| Humidity | -0.0605 | 0.0161 | -0.0948 | -0.0288 |
| Max_Wind | -0.0930 | 0.0598 | -0.2132 | 0.0178 |
| linearTrend | 28.7428 | 3.7136 | 23.1412 | 37.678 |

**Link function: log**
**Distribution family: Poisson**
**Number of coefficients: 7**
**Log-likelihood: -735.4298**
**AIC: 1484.86**
**BIC: 1496.504**
**QIC: 1484.86**

The model diagnostics are presented in Figure 4. Autocorrelation function of residuals, shown in Figure 4 (top left), does not exhibit any serial correlation or seasonality which has not been taken into account by the model. Figure 4 (bottom left) points to a probability integral transform (PIT) histogram which appears to approach uniformity. Figure 4 (top right) indicates the overall behavior of the data set. Figure 4 (bottom right) shows a marginal calibration plot with minor fluctuations about zero. Hence the probabilistic calibration of the Poisson model is sufficient.

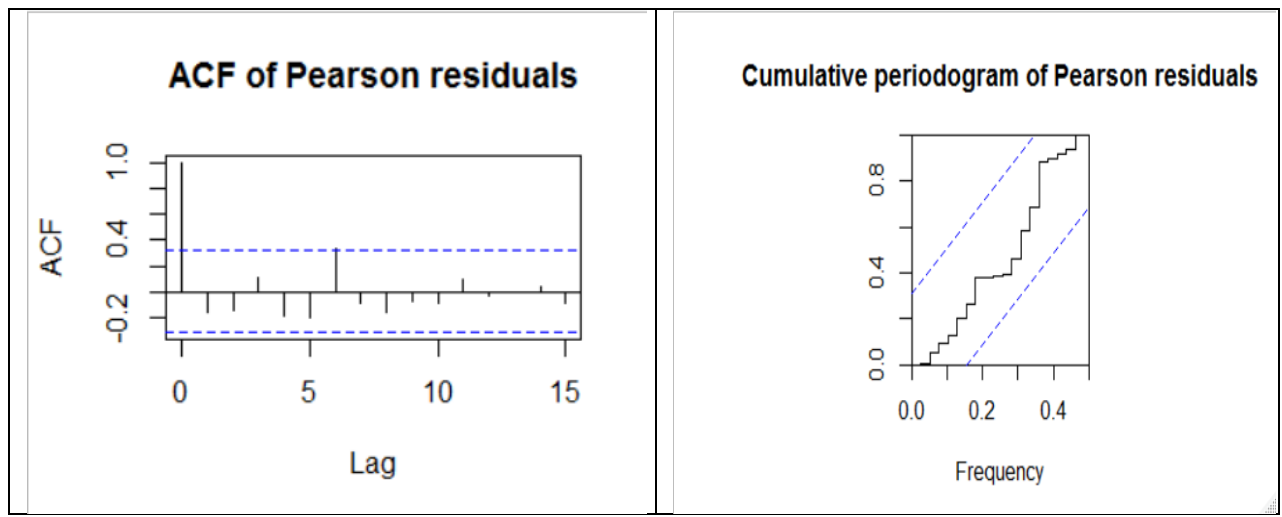

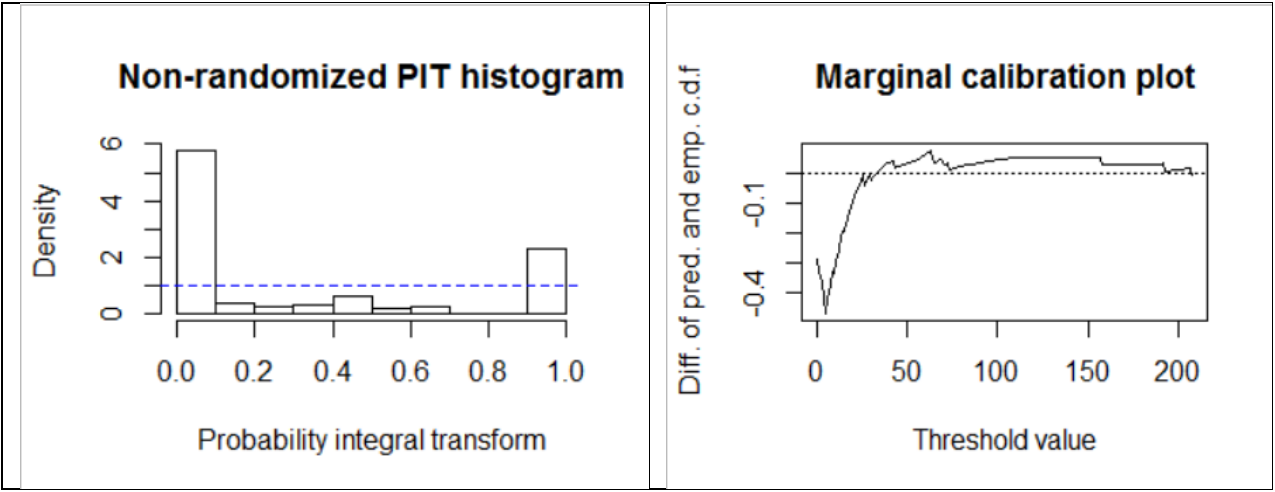

**Figure 4: Diagnostic plots after model fitting to the data**

### 3.3 Model validation

Time series of the actual daily new cases of COVID-19 used for model fitting and predicted cases based on the fitted model are shown in Figure 5. It is observed that the overall trend of the two curves is similar, and the values themselves are very close in some cases. This is an indication of a good model.

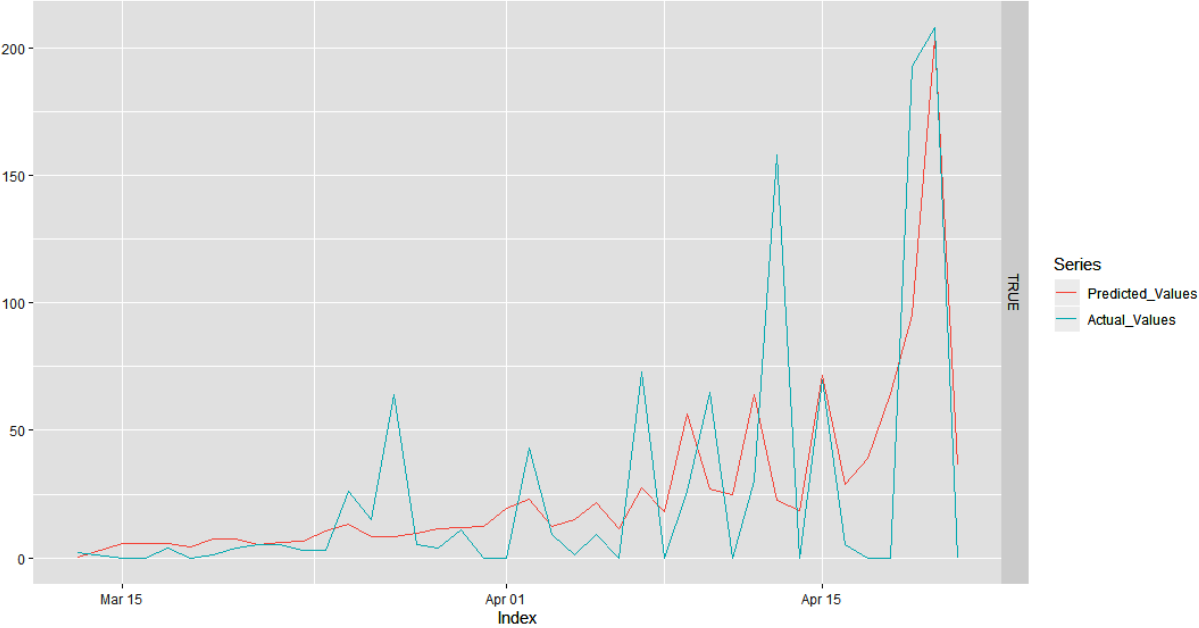

**Figure 5: In-sample model validation**

*3.3 Intervention analysis*

Here, we test whether there was an abrupt shift in the number of new cases of COVID-19 occurring when the partial lockdown was introduced on March 30, 2020. The approximate score test was applied. With a *p*-value of 0.0164 (Table 3), the null hypothesis of no significant effect of the specific policy intervention of partial lockdown is rejected at a 5% significance level. Hence, partial lockdown of the two largest cities and their environs in Ghana significantly influenced the daily confirmed cases of COVID-19.

**Table 3: Chi-square test of the effect of partial lockdown on new COVID-19 cases in Ghana**

| Chi-Square Statistic | Degrees of freedom | p-value |
| --- | --- | --- |
| 8.2247 | 2 | 0.0164 |

## 4 Discussion

In this study, we explored the possible effects of weather and government policy intervention on the daily confirmed cases of COVID-19. Specifically, the study considered the policy of partial lockdown of the two major cities (Accra and Kumasi) and their environs. The generalized linear model (GLM) framework for time series of counts by [17] was adopted for model fitting and analysis, which allows for regressing on past observations of the response variable and covariates.

The results indicated that daily confirmed cases of COVID-19 in Ghana correlated negatively with maximum temperature and relative humidity. Previous studies on SARS-CoV and MERS-CoV also highlighted temperature as an important factor in the survival and transmission of such coronaviruses [22]–[26].

Recent studies have also underscored the importance of weather on the transmission of COVID-19 in particular. For instance, using a Modified Susceptible-Exposed-Infectious-Recovered (M-SEIR) model, [27] confirmed that transmission rate decreased with the increase of temperature, leading to further decrease of infection rate and outbreak scale in 31 provincial-level regions in mainland China. [28] also in a study suggested that for northern hemisphere countries, the rate of transmission of COVID-19 cases would significantly decrease as a result of warmer weather. [29] in a study stated that local weather condition with low temperature, mild diurnal temperature range and low humidity likely favor the transmission of COVID-19 cases in China. [30] in a study, analyzed the association between COVID-19 and climate indicators in New York City, USA. The authors used a secondary published data from New York city health services and National weather service, USA. The climate indicators included in the study were average temperature, minimum temperature, maximum temperature, rainfall, average humidity, wind speed, and air quality. Using Kendall and Spearman rank correlation tests for the data analysis, [30] suggested that that average temperature, minimum temperature, and air quality were significantly associated with COVID-19. Finally, using a global data on COVID-19 reported cases until 29th February 2020 and climate temperature data, [31] suggested that higher average temperature was strongly associated with lower COVID-19 incidence for temperatures of 1°C and higher.

Our results further indicated the significance of maximum temperature, relative humidity and precipitation in predicting daily confirmed cases of COVID-19 in Ghana. This is consistent with the results of [32], who implemented a restricted cubic spline function and generalized linear mixture model to analyze the relationships between temperature and COVID-19 transmission in 429 cities in the world and suggested that to a certain extent, temperature could significantly change COVID-19 transmission, and there might be a best temperature for the viral transmission, which may partly explain why it first broke out in Wuhan, China. [33] also investigated the influence of air temperature and relative humidity on the transmission of COVID-19 in 100 Chinese cities and suggested that, under a linear regression framework, high temperature and high humidity significantly reduce the transmission of COVID-19.

Additionally, results of the intervention analysis conducted in this study indicated that the null hypothesis of no significant effect of the specific policy intervention of partial lockdown should be rejected (p-value=0.0164) at a 5% level of significance. This finding underscores the significance of the partial lockdown of the two largest cities (Accra and Kumasi) and their environs in controlling the spread of the disease in Ghana. A number of studies [14]–[16] have been conducted on the effects of lockdown on the number of new confirmed cases and/or on deaths. All of which reveal that lockdown serves an important intervention for preventing and minimizing the impact of an outbreak.

## 5    Conclusion

In this paper, we investigated the effects of weather and government policy intervention on the daily confirmed cases of COVID-19 in Ghana. The results indicated that while daily confirmed cases of COVID-19 correlated negatively with maximum temperature and humidity, it rather correlated positively with precipitation and maximum wind. Also, maximum temperature, humidity and precipitation were significant predictors of daily confirmed cases of COVID-19 in Ghana. Furthermore, results of the intervention analysis revealed that the specific government policy of partial lockdown of the two largest cities (Accra and Kumasi) and their surroundings significantly influenced the daily confirmed cases of COVID-19, which provides useful implications for policy makers and the public. For instance, knowing that precipitation and the daily confirmed cases of COVID-19 are positively correlated could help public health responses by informing key preparations as we move into the rainy season.